\documentclass[11pt]{article}
\usepackage{moriond,epsfig}
\usepackage{slashed}

\def\Journal#1#2#3#4{{#1} {\bf #2}, #3 (#4)}

\def\be{\begin{equation}}
\def\ee{\end{equation}}
\def\bea{\begin{eqnarray}}
\def\eea{\end{eqnarray}}

\begin{document}
\vspace*{4cm}
\title{NON-SUSY SEARCHES AT ATLAS}

\author{ E.N. Thompson \\ (on behalf of the ATLAS Collaboration)}

\address{Department of Physics\\ 
1126 Lederle Graduate Research Tower (LGRT)\\
University of Massachusetts\\
Amherst, MA 01003
}

\maketitle\abstracts{
The ATLAS detector has begun the search for new physics beyond the Standard Model (BSM) with an integrated luminosity of $\int Ldt\simeq45$~pb$^{-1}$
of data collected in 2010. After no significant evidence of new physics was found in the data, limits on possible 
signatures have been set, many of which are already more stringent than previous measurements. These proceedings 
review recent limits obtained on various BSM models, 
including excited quarks, axigluons, contact interactions, quantum black holes, heavy gauge bosons ($W'$, $Z'$), gravitons, fourth-generation quarks
and leptoquarks.
}

\section{Introduction}

Over the past few decades, experimental results have consistently agreed with the observable 
expectations of the Standard Model (SM); 
%yet the Higgs mechanism, 
yet the cause of Electroweak Symmetry Breaking (EWSB), 
necessary to give mass to the $W^{\pm}$ and $Z^0$ bosons, remains unconfirmed.
%   While the 
% Higgs mechanism tends to be the most popular theoretical explanation for EWSB, 
% it suffers from the ``Hierarchy Problem'' (quadratic divergences of
% radiative corrections to the Higgs mass). 
In addition, the SM contains 21 arbitrary 
parameters, and is unable to account for the number of quark/lepton families, 
matter-antimatter asymmetry, or gravity.  
Many theories beyond the Standard Model (BSM) 
have been developed to address limitations of the SM.
The ATLAS detector~\cite{detector}, located on the Large Hadron Collider (LHC) ring at CERN
near Geneva, Switzerland, was especially designed to measure high-momentum particles in anticipation of new physics
discoveries at the electroweak scale ($\sim$1~TeV).
Signatures in the detector span a large range of final-state objects,
which include electrons, photons, jets, missing transverse energy ($\slashed{E}_T$) and muons.
% ATLAS is comprised of multiple subdetectors, each with the capacity to withstand large particle 
% fluxes and measure these high-momentum particles with great precision.

In the early months of data-taking, understanding the performance of the detector was crucial while searching 
for new physics processes occurring at a much higher invariant mass than available at previous
experiments.  ATLAS collected a total of $\simeq45$~pb$^{-1}$ of integrated luminosity in 2010
at a center of mass energy $\sqrt{s}=7$~TeV, and quickly `rediscovered' many SM physics processes, setting the stage
for new physics discovery. These proceedings discuss the new physics searches performed at ATLAS according to their 
final-state signatures, as well as their related detector performance issues at high momentum.

%\section{New Physics Searches}

Slightly different datasets were used for each analysis due to varying trigger requirements and data quality conditions of the
subdetectors used to measure the final-state objects. The uncertainty on the luminosity measurement for
the analyses below was 11\%.

\section{Searches in the dijet final state}

The 2$\rightarrow$2 scattering processes as described by QCD in the Standard Model have been well studied, and any 
deviation from expected behavior of dijet processes would indicate new physics.  Searches were performed in both the 
dijet invariant mass spectrum~\cite{DijetsPaper}, given by $$m_{jj} = \sqrt{(E_{j_1}+E_{j_2})^2 - (\vec{p}_{j_1}+\vec{p}_{j_2})^2},$$
as well as in the angular distribution $F_\chi(m_{jj})$ of dijets, where 
$$F_\chi\left(\frac{[m_{jj}^{min}+m_{jj}^{max}]}{2}\right)\equiv\frac{N_{events}(|y^*|<0.6,~m_{jj}^{min},~m_{jj}^{max})}
{N_{events}(|y^*|<1.7,~m_{jj}^{min},~m_{jj}^{max})}~.$$ 
Here, $N_{events}$ are the number of dijet events observed within the rest-frame rapidity $y^*$ and invariant mass ranges specified.
Models studied appearing as a resonance in the dijet mass 
spectrum include exited quarks~\cite{excitedq1} and axigluons~\cite{axigluons1}. 
Other signals could also appear as a non-resonant excess of 
events above the dijet invariant mass distribution, such as in $qqqq$ contact interactions~\cite{CI1} or 
quantum black hole (QBH)~\cite{QBH} models.
More sensitivity to new physics may be gained by using the angular distribution, as QCD dijets are more 
central in the detector, while new physics signatures are more isotropic in nature. An analysis using the $F_\chi(m_{jj})$ distribution 
also benefits from less sensitivity to the absolute jet energy scale (JES) which is the largest systematic 
uncertainty for high-energy jets.   

% After selecting data runs with the best possible detector conditions in the ID and calorimeters, 36~pb$^{-1}$ 
% of integrated luminosity was used in the dijet analsyes. Events were selected with at least one collision vertex,
% and with $p_T^{j}>$ 60 GeV (30 GeV) for the leading $p_T$ (secondary) jet. 

Using the calorimeter trigger and requiring high quality data in the Inner Detector (ID) and calorimeters, 36~pb$^{-1}$ 
of integrated luminosity was used in the dijet analyses. Each of the two jets in the
event were required to pass quality criteria ensuring that the energy deposition in the 
calorimeters was in-time. 
%Pooly reconstructed jets with $p_T^{j}>$ 15 GeV which could be identified as one of the 
%two primary jets were vetoed~\cite{jetstuff}. 
For the dijet resonance search, the two selected jets were additionally required to have a pseudorapidity $|\eta_j|<2.5$, $|\Delta\eta_{jj}|<1.3$ between
them, and leading jet $p_T^{j_1}>$~150 GeV, leaving 98,651 events with $m_{jj}>$500 GeV passing all selection.  In the angular distribution analysis,
the additional selection required dijets to satisfy tighter rapidity ranges, with 71,402 events in data after all selection.
In both the resonance and angular distribution searches, the data were found to be consistent
with SM expectations.  Limits were set on the models mentioned above using a modified frequentist (CL$_{s+b}$)
approach (for $F_\chi(m_{jj})$ searches) as well as a Bayesian credibility interval approach (for $m_{jj}$ searches).
The 95\% C.L. lower limits are summarized in Table~\ref{tab:dijetlimits}.

\begin{table}[ht]
  \begin{center}
    \caption{95\% C.L. lower limits on various dijet physics signatures. Units are in TeV.  The limit for QBH is given
for number of dimensions $>$ 6, and the Contact Interaction limit is set on the scale of the new interaction $\Lambda$.\label{tab:dijetlimits}}
    \vspace{0.3em}
    \begin{tabular}{|l|cc|cc|}
     \hline
                  & \multicolumn{2}{|c}{$m_{jj}$} & \multicolumn{2}{c|}{$F_\chi(m_{jj})$} \\ 
    Model         & \multicolumn{1}{|c}{Expected} & \multicolumn{1}{c}{Observed} & \multicolumn{1}{c}{Expected} & \multicolumn{1}{c|}{Observed} \\  
     \hline
    Excited Quark & 2.07 & 2.15 & 2.12 & 2.64 \\
    QBH           & 3.64 & 3.67 & 3.49 & 3.78 \\
    Axigluon      & 2.01 & 2.10 & -    & -    \\
    Contact Interaction & - & - & 5.72 & 9.51 \\
     \hline
    \end{tabular}
  \end{center}
\end{table}

\section{Searches in the charged dilepton, lepton with $\slashed{E}_{T}$ and diphoton final states}

Some extensions to the SM predict massive gauge bosons ($W'$, $Z'$) above 1~TeV. In the Sequential Standard Model (SSM)~\cite{ZprimeSSM}, the couplings 
of the fermions to the $W'$ or $Z'$ are the same as for the SM $W$ and $Z$ bosons. In another string-theory-inspired model~\cite{ZprimeE6}, an
$E_6$ gauge group symmetry-breaking leads to 6 different $Z'$ states: $Z'_\psi$, $Z'_N$, $Z'_I$, $Z'_S$, $Z'_\eta$ and $Z'_\chi$.
Additionally, models with excited states $W^*$ and $Z^*$~\cite{ZprimeStar} were also considered, as they have different kinematic distributions 
than those of the $W'$ and $Z'$.
Searches were performed in the dilepton channel~\cite{ZprimePaper} ($Z'\rightarrow e^+e^-$ and $Z'\rightarrow \mu^+\mu^-$) as well 
as the lepton plus $\slashed{E}_T$ channel~\cite{WprimePaper} ($W'\rightarrow e\nu_e$ and $W'\rightarrow \mu\nu_\mu$). Similarly to the 
dijets searches, the invariant mass of the dileptons $m_{ll}$ was used in order to search for $Z'$ resonances. However, as the neutrino appears only
as missing energy in the detector, only the transverse component of the mass $m_T = \sqrt{p_T^l\slashed{E}_T(1-\cos\phi_{l\nu})}$
could be used in $W'$ searches. The total integrated luminosity for the $W'$ searches was 36~pb$^{-1}$, while the searches
for $Z'\rightarrow e^+e^-$ and $Z'\rightarrow \mu^+\mu^-$ used 39~pb$^{-1}$ and 42~pb$^{-1}$, respectively.

The underlying physics process is the same regardless of the participant leptons, however, the systematic uncertainties 
entering the analyses are different for the electron and muon channels and are handled separately.  
Higher-order QCD processes are less understood at high mass, and so the background 
to $W'\rightarrow e\nu_e$, $W'\rightarrow \mu\nu_\mu$ and $Z'\rightarrow e^+e^-$ signals arising from jets faking 
electrons or mis-measuring $\slashed{E}_T$ was estimated based on control regions in the data. In the muon channel, the largest uncertainty
came from muon momentum resolution, mainly due to misaligned Muon Spectrometer (MS) chambers which were not modeled in simulation. This was 
reduced by requiring that muons use combined measurements in both the ID and MS, and also pass through regions
of the MS with higher geometric acceptance and better-known alignment.
In both channels of the $Z'$ search, the total systematic uncertainty was reduced by normalizing the data 
to the simulation in the control region $70<m_{ll}<120$~GeV.

After all selection criteria, 31 (16) events were observed in the $e\nu$ ($\mu\nu$) 
channel having $m_{T}\nobreak>\nobreak500$~GeV, and 66 (38) events observed in the $ee$ ($\mu\mu$) channel in
the signal region $m_{ll}>150$~GeV. 
In both cases, the data agreed with SM expectations. Limits on the new physics cross sections were set using a 
modified frequentist approach for the $W'$ searches and a Bayesian approach for the $Z'$ searches.
The 95\% C.L. lower limits after combining the electron and muon channels are summarized in Table~\ref{tab:WZprimeLimits}.
\begin{table}[ht]
  \begin{center}
    \caption{95\% C.L. lower limits on various dilepton and lepton+$\slashed{E}_T$ physics
            signatures after electron and muon channels were combined. Units are in TeV. \label{tab:WZprimeLimits}}
    \vspace{0.3em}
    \begin{tabular}{|l|cccccccccc|}
     \hline
     Model & $W'$ & $W^*$ & $Z'$ & $Z^*$ & $Z'_\psi$ & $Z'_N$ & $Z'_I$ & $Z'_S$ & $Z'_\eta$ & $Z'_\chi$ \\
     \hline
     Expected & 1.450 & 1.320 & 1.088 & 1.185 & 0.837 & 0.860 & 0.922 & 0.945 & 0.866 & 0.965 \\
     Observed & 1.490 & 1.350 & 1.048 & 1.152 & 0.738 & 0.763 & 0.842 & 0.871 & 0.771 & 0.900 \\
     \hline
    \end{tabular}
  \end{center}
\end{table}

Searches were also performed in the diphoton channel~\cite{DiphotonPaper} where a Kaluza-Klein graviton resonance is predicted by the Randall-Sundrum 
model~\cite{RSgraviton}. In this model an extra spacial dimension, characterized by its curvature $k$ and compactification radius $r_c$, 
reduces the Plank scale $\bar{M}_{pl}$ down to the TeV scale.  Because photons were reconstructed in the EM calorimeter, they
had similar detector-related uncertainties as for the dielectron channel. No excess above the predicted diphoton invariant mass spectrum
was found in 36~pb$^{-1}$, and limits on various combinations of the graviton mass ($m_{\gamma\gamma}$) and couplings ($k/\bar{M}_{pl}$) 
were set using a modified frequentist approach. This resulted in a 95\% C.L. lower limit of 545~(920)~GeV on the RS graviton mass 
with coupling $k/\bar{M}_{pl}$ = 0.02 (0.1).

\section{Searches with leptons and jets}

Finally, other models predict multiple-object final-states, such as fourth generation quarks~($Q_4$)~\cite{4thQuarkPaper}
and those involving new particles carrying both lepton and baryon number (``leptoquarks'')~\cite{LQPaper}.
In the decay of pairs of fourth generation quarks ($Q_4Q_4\rightarrow W^+jW^-j\rightarrow l^+\nu jl^-\nu j$), boosted $W$ bosons 
result in a final-state in which lepton-jet pairs are more collinear than in SM $W$ pair production. 
A transverse ``collinear mass'' was constructed from lepton-jet pairs after assuming that the neutrinos were the only contributors to $\slashed{E}_T$ and
by choosing $|\Delta\eta(\nu,l)|$, $|\Delta\phi(\nu,l)|$ and jet assignments which minimized the difference between the 
two collinear masses in the event. Together, the scalar sum of the transverse energy in the event and the collinear mass
was used as a discriminant to separate new physics from the dominant top quark pair production background. In
37~pb$^{-1}$, no evidence for fourth generation quarks was found, and a limit of $M_{Q_4}>270$~GeV (95\% C.L.) was set
using a Feldman-Cousins approach. This was the first dilepton search for an up-type fourth generation quark, as well
as the first search for fourth generation quarks performed at the LHC.

Searches for leptoquarks were performed %in $ll\nu j$ and $lljj$ final states 
using 35~pb$^{-1}$ of data in
both the first ($eejj$ and $e\nu jj$) and second ($\mu\mu jj$ and $\mu\nu jj$) generation channels.
The sum of transverse energy in the event $S_T$ $\equiv\sqrt{p_T^{l_1}+p_T^{l_2}+p_T^{j_1}+p_T^{j_2}}$ 
was used to look for the presence of a new physics signal in the $llqq$ channel. Here, in order to
reject background events arising mainly $Z$ with jets, $M_{ll}$ was found by computing the invariant mass of pairs of leptons.
%The  $M_{LQ}$ was found by computing the invariant mass of pairs of leptons and jets. 
In the $ll\nu j$ channel, however, because the neutrino could not be fully reconstructed, only the transverse mass was used to
reject the dominant $W$ plus jets background (defined similarly as for the lepton-neutrino pair in the $W'$ search). 
Choosing lepton-jet and jet-neutrino pairs which minimize the difference between the invariant mass of the lepton-jet $M_{LQ}$ and 
transverse mass of the jet-neutrino $M_{LQ}^T$, the average 
$\bar{M}_{LQ}$ between them was used as a discriminant in the $l\nu qq$ channel. The data were found to be consistent with 
SM expectations, and using a modified frequentist approach, 95\% C.L limits on the leptoquark mass were set 
as a function of the branching fraction $\beta$ for a single leptoquark to decay to a charged lepton + jet. For 
first generation leptoquarks, lower limits of $M>376$~(319)~GeV for $\beta$=1.0 (0.5) were found. In the second
generation, the lower limits found were $M>422$~(362)~GeV for $\beta$=1.0 (0.5).

\section{Summary}

ATLAS has shown that it is able to push the energy frontier to the TeV scale and search for a variety
of BSM signatures in the earliest stages of LHC data, in many cases setting the world's best limits. 
The knowledge gained from the first round of analyses presented at Rencontres de Moriond has paved the way 
for future work, and ATLAS eagerly awaits more data in the hopes of discovering new physics at the LHC.

%\section*{Acknowledgments}


\section*{References}
\begin{thebibliography}{99}

\bibitem{detector}The ATLAS Collaboration, G. Aad et al.,
\textit{The {ATLAS} {E}xperiment at the {CERN} {L}arge {H}adron {C}ollider,}~\Journal{JINST}{3}{S08003}{2008}

\bibitem{DijetsPaper} The ATLAS Collaboration, G. Aad et al., \textit{Search for New Physics in Dijet Mass and Angular 
Distributions in pp Collisions at $\sqrt{s}$=7 TeV Measured with the ATLAS Detector,}~\Journal{New J. Phys.}{13}{053044}{2011}

\bibitem{excitedq1}U. Baur, I.Hinchliffe, and D. Zeppenfeld,~\textit{Excited Quark Production at Hadron Colliders,}~
\Journal{Int. J. Mod. Phys}{A2}{1285}{1987}\\
U. Baur, M. Spira, and P.M. Zerwas,~\textit{Excited-quark and -lepton production at hadron colliders,}~\Journal{Phys. Rev.}{D42}{815}{1990}

\bibitem{axigluons1}P. Frampton and S. Glashow,~\textit{Chiral Color: An Alternative to the Standard Model,}~\Journal{Phys. Lett.}{B190}{157}{1987}\\
%P.~Frampton and S.~Glashow~~\Journal{Phys. Rev. Lett.}{58}{2168}{1987};
J. Bagger, C. Schmidt, and S. King,~\textit{Axigluon Production in Hadronic Collisions,}~\Journal{Phys. Rev.}{D37}{1188}{1988}

\bibitem{CI1}E. Eichten, I.Hinchliffe, K.D. Lane and C. Quigg~\textit{Supercollider Physics}~\Journal{Rev. Mod. Phys.}{56}{579}{1984}\\
%E.~Eichten, I.Hinchliffe, K.D. Lane and C. Quigg~~\Journal{Rev. Mod. Phys.}{58}{1065}{1986};
P.~Chiappetta and M. Perrottet,~\textit{Possible bounds on compositeness from inclusive one 
jet production in large hadron colliders,}~\Journal{Phys. Lett.}{B253}{489}{1991}

\bibitem{QBH}P. Meade and L. Randall,~\textit{Black Holes and Quantum Gravity at the LHC,}~\Journal{JHEP05}{2008}{003}{2008}

% \bibitem{jetstuff}ATLAS Collaboration,~\textit{Data-Quality Requirements and Event Cleaning for Jets and Missing $E_T$ Reconstruction 
% with the ATLAS Detector in $pp$ Collisions at $\sqrt{s}$=7 TeV,}~ATLAS-CONF-2010-038 (http://cdsweb.cern.ch/record/1277678)

\bibitem{ZprimeSSM}P. Langacker,~\textit{The Physics of Heavy Z' Gauge Bosons,}~\Journal{Rev. Mod. Phys.}{81}{1199}{2009} 

\bibitem{ZprimeE6}D. London and J. L. Rosner,~\textit{Extra Gauge Bosons in E$_6$,}~\Journal{Phys. Rev.}{D34}{1530}{1986} 

\bibitem{ZprimeStar}M. Chizhov, V. Bednyakov, and J. Budagov,~\textit{Proposal for chiral-boson search at LHC 
via their unique new signature,}~\Journal{Physics of Atomic Nuclei}{71}{2096}{2008} 

\bibitem{ZprimePaper} The ATLAS Collaboration,~\textit{Search for high mass dilepton resonances in $pp$ collisions 
at $\sqrt{s}$=7 TeV with the ATLAS experiment,} arXiv:1103.6218,~accepted by Phys. Lett. B (2011)

\bibitem{WprimePaper} The ATLAS Collaboration,~\textit{Search for high-mass states with one lepton plus missing transverse momentum 
in proton-proton collisions at $\sqrt{s}$ = 7 TeV with the ATLAS detector,}~accepted by Phys. Lett. B (2011)

\bibitem{DiphotonPaper} The ATLAS Collaboration,~\textit{A Search for High Mass Diphoton Resonances 
in the Context of the Randall-Sundrum Model in $\sqrt{s}$ = 7 TeV $pp$ Collisions,}~ATLAS-CONF-2011-044 (http://cdsweb.cern.ch/record/1338573)

\bibitem{RSgraviton}L. Randall and R. Sundrum,~\textit{Large Mass Hierarchy from a Small Extra Dimension,}~\Journal{Phys. Rev. Lett.}{83}{3370}{1999}

\bibitem{4thQuarkPaper} The ATLAS Collaboration,~\textit{Search for Fourth Generation Quarks Decaying 
to $WqWq\rightarrow ll \nu \nu q q$ in $pp$ collisions at $\sqrt{s}$=7 TeV with the ATLAS Detector,}~ATLAS-CONF-2011-022
 (http://cdsweb.cern.ch/record/1336751)

\bibitem{LQPaper} The ATLAS Collaboration,~\textit{Search for pair production of first or second generation leptoquarks 
in $pp$ collisions at $\sqrt{s}$=7 TeV using the ATLAS detector at the LHC,}~arxiv:1104.4481, submitted to Phys Rev D (2011)


\end{thebibliography}
\end{document}